\documentstyle[aps,psfig,multicol]{revtex}
\begin{document}
\topmargin -15mm

\title{Electron correlation effects and magnetic ordering at the Gd(0001) surface
}

\author{A.B. Shick, W.E. Pickett and C. S. Fadley}

\address{Department of Physics, University of California, Davis, CA 95616}

\maketitle

\begin{abstract}
Effects of electron correlation
on the electronic structure and magnetic properties of
the Gd(0001) surface are investigated using of the full-potential linearized
augmented plane wave implementation of correlated band theory (``LDA+U").
The use of LDA+U instead of LDA (local density approximation) total energy calculations
produces the correct ferromagnetic ground state
for both bulk Gd and the Gd surface.
Surface strain relaxation leads to an 90 \% enhancement of the interlayer
surface-to-bulk effective exchange coupling. Application of a Landau-Ginzburg type theory
yields a 30 \% enhancement of the Curie temperature at the surface, in very good
agreement with the experiment.\\
\end{abstract}

\begin{multicols}{2}
Many of the magnetic properties of Gd metal are well understood \cite{Macintosh}.
The half-filled 4f-shell $(S= \frac{7}{2}, L= 0)$ of Gd leads to a formation
of a well localized spin-only magnetic moment. These localized spin moments
couple through an RKKY-type exchange interaction to form a ferromagnetic (FM)
Heisenberg system with a bulk Curie temperature ($T^b_c$) of 293 K \cite{Macintosh}. 
FM order polarizes the conduction electrons and leads to a total magnetic moment
of 7.63 $\mu_B$ / Gd atom \cite{Roeland}.

However, in spite of relatively simple bulk magnetic behavior, the magnetism of the
Gd surface is rather unusual \cite{Dowben}.
The results of different spectroscopic measurements suggest a significant enhancement of the
surface Curie temperature ($T^s_c$) for Gd(0001). Gd is thus one of only three ferromagnets
(including Tb and FeNi$_3$) for which such an increased $T^s_c$ has been observed \cite{Rau}.
After a first observation of this effect in Gd by Rau {\it et al.} \cite{Rau},  
Weller {\it et al.} verified it for 400 $\AA$ thick Gd(0001) films grown on a W(110) substrate,
by comparing spin-polarized low energy electron diffraction (LEED)
and the magneto-optic Kerr effect measurements \cite{Weller1}, and further suggested
a possible anti-ferromagnetic
(AFM) alignment of the surface layer(s) with respect
to the bulk FM Gd.
Further investigations with spin-polarized valence and core photoemission spectroscopy (PES)  \cite{Erskine}
did not confirm an existence of this
surface AFM coupling.
Instead, the in-plane component of surface layer magnetization was observed to be
parallel to the bulk \cite{Erskine,Weller2,Li}, althrough the possibility of canted or mixed in-plane and
out-of-plane surface magnetic ordering was suggested.
Very recent spin-polarized photoelectron diffraction experiments \cite{Tober}
for bulklike $\approx$ 300  $\AA$ thick epitaxial Gd/W(110) films clearly indicate
temperature dependent core level spin asymmetries well above the bulk $T^b_c$, also suggesting
surface enhancement of $T^s_c$ of as much as 85 K. However, to date, there has been
no quantitative theoretical explanation for this enhanced $T^s_c$.

The aim of this paper is to 
show that first principles calculations which account for both electronic
correlations of the 4f electrons and the relaxation of the surface atomic positions can
provide such a quantitative description of the
electronic and magnetic structure of the Gd(0001) surface.
Using the results of total energy calculations and the Landau-Ginzburg
model we show that there is an increase of $T_c$ at the Gd surface due to
an enhanced surface-to-bulk effective exchange coupling that is in turn caused by the surface
structure relaxation. 
We emphasize the substantial role
of electron correlations for the Gd f-electrons to obtain a correct FM ground state for
both bulk Gd and the Gd surface. 
We also emphasize the role of structure relaxation as a ``driving force" for the $T_c$
enhancement.

{\it 1.Difficulties of local (spin) density approximation (LDA) and generalized gradient approximation (GGA)
and limitations of a core treatment for Gd 4f electrons.}
Since the pioneering work of Dimmock and Freeman \cite{Art1} there have been other
attempts to describe the 4f states of Gd in terms
of the localized ``4f-core'' electron model \cite{Eschrig}
(in which the 4f states of Gd
are treated as a part of fully localized atomic core). 
Singh \cite{Singh} performed a detailed analysis of
the limitations of this ``4f-core" model
and  
achieved very good agreement with experiment for the ground state lattice constant and
the magnetic moment of FM bulk Gd by using a ``4f-band" model and LDA, 
but he did not consider a possible AFM
phase.
Recent full potential linear muffin-tin orbitals (LMTO) calculations \cite{Harmon,Eriksson}
show that both the LDA and the GGA yield an AFM phase that
is lower in energy than the FM phase.
This problem is solved by employing the  LDA+U \cite{lixt} method to treat
the electron correlations for the 4f-electrons of Gd. 
It was demonstrated \cite{Harmon} that 
the use of LDA+U \cite{lixt} instead of LDA yields
a correct FM ground state and also
provides 4f electron binding energies in good agreement
with experiment.
Recently we confirmed quantitatively the conclusions of Ref. \cite{Harmon}
for bulk Gd using the LDA+U total energy functional with the
full-potential linearized augmented plane wave (FLAPW) method \cite{ldau}.

The first total energy FLAPW calculations of the magnetic ordering
at the Gd surface \cite{Wu} using the 4f-core model
reported an AFM coupling of the surface layer with respect to
the bulk, in agreement with early interpretation of experimental data \cite{Weller1}.
The full potential LMTO calculations using the 4f-core model
\cite{Eriksson} did not reproduce the results of Ref. \cite{Wu} and yielded 
FM coupling between the surface and bulk magnetization, in agreement with
the most recent experiments \cite{Erskine,Weller2,Li}.
It is thus clear that:
(i) the 4f-band model with LDA fails to
account for the correct magnetic ordering for bulk Gd,
and the use of GGA instead of LDA does not improve the situation;
(ii) two LDA   4f-core model calculations with LDA yield 
conflicting results (\cite{Eriksson} and  \cite{Wu})  for the magnetic ground state
at the Gd surface.

Since the LDA+U method works well to describe the electronic structure
and magnetic ground state for bulk Gd, we decided to apply it in electronic
structure calculations and total energy determinations of the magnetic behavior
of the Gd surface. 

{\it 2.Computational results.}
The FLAPW method \cite{flapw}
is employed to perform scalar-relativistic self-consistent film calculations for
Gd. 
The fully relativistic self-consistent version \cite{flapwso} of this method
is then used to perform the final LDA calculations.
The LDA+U calculations are based on the 
scalar-relativistic version of the FLAPW method \cite{ldau}. 
The literature values \cite{Harmon} of the on-site repulsion $U \; = \; 6.7 $ eV
and exchange $J \; = \; 0.7$ eV were used in the calculations. 

For the Gd(0001) surface, we choose the isolated slab model based on 7-layer Gd film (with z-reflection symmetry)
and in the first set of calculations use the bulk lattice constant (3.634 \AA) and c/a ratio (1.587) \cite{struc}.
Here,  32 special k-points \cite{BZ} in the irreducible 1/3 part of the 2D BZ \cite{symmetry}
were used, with Gaussian smearing for k-points weighting.
The ``muffin-tin'' radius values of $R_{MT} \; = \; 3.2 \; a.u.$ and $R_{MT} \times K_{max} \; = \; 9.6$
(where, $K_{max}$ is the cut-off for LAPW basis set)  were used.  

{\it 2A.LDA results.}
The spin magnetic moments for a Gd film with its surface layer magnetically coupled
parallel ($\uparrow \uparrow$) and antiparallel ($\downarrow \uparrow$) to
the FM bulk
resulting from  the scalar-relativistic LDA calculations
show that the magnetically active
f states are almost fully polarized (with the magnetic moment 6.88 $\mu_B$ in the bulk and 6.82 $\mu_B$ at the surface)
and induced spin polarization of $\approx$ 0.5 $\mu_B$/atom of conduction
electrons (mainly d states). This magnetic coupling
is a result of intraatomic inter-band exchange interaction between conduction band and 
localized f electrons as incorporated in the s-f exchange model \cite{Vonsovski} 
and can be understood to be due to
a positive inter-band d-f exchange coupling \cite{Nolting}.
There is a slight decrease of the 4f magnetic moment at the surface layer
due to an increase of minority spin 4f occupation.

Starting from the results of scalar-relativistic calculations, we then performed self-consistent
relativistic LDA calculations for a Gd-film, assuming [0001] spin axis direction.
The spin moments are slightly decreased for f states in comparison with
scalar-relativistic calculations, due to an increase of minority spin
occupation of the 4f states.
The small spin-orbit induced orbital magnetic moments (0.14 $\mu_B$ for the bulk and 0.33 $\mu_B$ for the surface
atoms) are mainly due to the 4f minority spin contribution. A parallel coupling
between spin and orbital moments for 4f states is consistent with the 3rd Hund rule.
The orbital moments  from 5d states
are about $0.02 \mu_B$ per Gd atom and coupled anti-parallel to the spin moments,
again consistent with the 3rd Hund rule.
The values of total magnetic moment (the sum of spin and orbital moments) are close to
the values of spin moment from scalar-relativistic calculations.

The total energy difference $\Delta E_{({\downarrow \uparrow} - {\uparrow \uparrow})}$ 
between the two surface magnetic configurations $\uparrow \uparrow$ and $\downarrow \uparrow$
defined above 
is positive (36 meV/atom in scalar-relativistic calculations) and
does not change appreciably when spin-orbit coupling is included (40 meV/atom).
The effect of the spin-orbit interaction is seen to be very small for the energetics of Gd
due to the fact that the 4f spin-majority band is fully occupied
and the 4f spin-minority band is almost empty. Therefore, the spin-orbit coupling 
does not affect the calculated values of magnetic and
total energy properties of Gd and does not assist in resolving the limitations of LDA.

{\it 2B.LDA+U results.}
The spin magnetic moments for a Gd film with surface layer
magnetically coupled parallel to the FM bulk 
resulting from  the scalar-relativistic LDA+U calculations are shown in Table \ref{tab1}.
There is a moderate enhancement of the magnetic moment of 4f states compared to LDA values ($\approx$ 0.1 $\mu_B$/atom)
due to an upward shift of 1.5 eV of minority spin 4f states.
There is practically no difference between surface and bulk f state magnetic
moments. The total magnetic moment at the surface layer is enhanced  
compared to the bulk mainly due to an increase of the d-state contribution.

The total energy difference
$\Delta E_{({\downarrow \uparrow} - {\uparrow \uparrow})}$
(71 meV/atom)
is  positive
and of the
same order of magnitude as the result of the 4f-core model \cite{Eriksson}.
The results of the present LDA+U calculations for both bulk
$\Delta E_{(AFM-FM)}$ (63 meV/atom \cite{ldau} - the difference in energies between
bulk AFM and FM spin configurations)
and surface
$\Delta E_{({\downarrow \uparrow} - {\uparrow \uparrow})}$ ($71 \; meV/atom$)
are in reasonable agreement
with the results of 4f-core model calculations (85 meV/atom for the bulk and 95 meV/atom for the surface) \cite{Eriksson}.
It shows that parallel coupling between surface and
bulk magnetization is energetically preferable and there is no anti-parallel surface-to-bulk magnetic
coupling for the Gd surface. This conclusion is consistent with  
experimental observations of the in-plane component of surface layer magnetization to be
parallel to the bulk \cite{Erskine,Weller2,Li}.

The electron density of states (DOS) for the case of (energetically preferred)
${\uparrow \uparrow}$ coupled surface layer are shown in Fig. 1.
There is a 4.5  eV downward shift of the majority spin
4f states and a 1.5 eV upward shift of minority spin 
4f states compared to the LDA calculation results.
This latter shift makes the minority spin 4f band practically empty and
corrects the fundamental error of LDA 4f-band model.
The exchange splitting of 4f states is enhanced in  LDA+U calculations by
the amount of the Hubbard U \cite{ldau} resulting in an 11 eV splitting of
majority and minority 4f states, close to the experimentally derived value (12 eV) \cite{Dowben}. 
The formation of a surface
state at the Gd surface clearly shows up as a peak of DOS
in the vicinity of Fermi level (cf. Fig. 1) due to majority d states.
From the DOS  it is clear that LDA+U
show strongly localized character of 4f-states for both bulk and surface.
However, the response of the 4f states to their environment does not allow them to be considered as a
true core states. 

In order to check numerical convergence of our results with respect to 
k-space integration, we increased the number of special k-points in the irreducible
part of 2D BZ \cite{symmetry} from 32 to 50 in self-consistent calculations and found
very little change in magnetic moment for both ${\downarrow \uparrow}$ and ${\uparrow \uparrow}$
magnetic configurations ($\leq 0.04 \mu_B$).
The calculated total energy difference 
$\Delta E_{({\downarrow \uparrow} - {\uparrow \uparrow})} = 72 \; meV$ per surface atom
agrees very well with its value of $71 \; meV$  for a smaller number of k-points.
  
{\it 3.Strain Relaxation and Magnetic Ordering at the Gd
Surface.}
LEED measurements \cite{LEED} show
that there is atomic structural relaxation near the Gd surface:
the interlayer distance between surface and sub-surface
Gd layers is about $2.6 \%$ smaller than its bulk value and the sub-surface-to-bulk
layer distance is about $1 \%$ bigger than its bulk value.
We have performed LDA+U calculations with the surface and sub-surface layers
(i) with interlayer distances taken from the experiment \cite{LEED} and (ii)
with interlayer relaxations taken to be half way between the experimental surface values
and the bulk values.
As in the case of an ideal Gd surface we have considered
two possible magnetic configurations 
(${\uparrow \uparrow}$, ${\downarrow \uparrow}$)
with the surface layer coupled parallel and antiparallel to the FM bulk Gd.
Here, 50 special k-points in the irreducible 1/3 part of the 2D BZ were used.

The surface relaxation  affects very little the values of the magnetic moment
in comparison with the ideal surface:
for both cases of ${\uparrow \uparrow}$ and 
${\downarrow \uparrow}$ coupled surface layer there is a
slight decrease in the values of the surface and sub-surface layers
magnetic moment due to the change of conduction band magnetization caused by
reduced interlayer distance.

There is, on the other hand, a surprisingly large enhancement of the magnetic coupling energy  
$\Delta E_{(\downarrow \uparrow - \uparrow \uparrow)}$ (cf., Table \ref{tab2}) due to the surface
relaxation: the energy difference increases by 90 \% in the comparison with the unrelaxed
structure.

As was already mentioned, there is considerable experimental evidence
of $T_c$ enhancement at the Gd surface.
Since 4f-magnetic moments are well localised and interact due to 
RKKY-type of exchange interactions,
the use of the Heisenberg-type Hamiltonian for the dependence of energy on spin configuration
is physically justified for Gd.
For the sake of simplicity, we neglect the long-range behavior
of exchange interactions for the Gd bulk and surface, assuming that significant
physics can be discussed in terms of nearest-neighbor (NN) interactions and neglect
anisotropy in exchange interaction
between a Gd atom and its six in-plane and six interplane
NN in the bulk and three interplane NN at the surface.
The spin Hamiltonian is then given by:

\begin{equation}
H = - B_0 \sum_i \hat{S}_i - \sum_{i} \sum_{\delta} J_{i,i+\delta} \hat{S}_i \hat{S}_{i + \delta}
\label{1}
\end{equation}
where, $B_0$ is an external field,
$J_{i,i+\delta}$ is an exchange coupling constant between the spin $i$
and its $\delta$ NN ($J^b$ in the bulk and $J^s$ at the surface) and $\hat{S}_i$ is a 
spin operator. We then apply ``molecular field'' theory \cite{Stanley} 
to Eq.(1). It leads to different molecular fields acting on the spin at the surface and
in the bulk, due to the different number of the interplane NN
(six in the bulk, three at the surface), and the difference between bulk and surface exchange coupling
constants $J_b$ and $J_s$. In the vicinity of the Curie temperature, when the value of the average
spin moment $<S(T)>$ is small, it is possible to introduce a Landau-Ginzburg type model for the temperature
dependence of $<S(T)>$ \cite{Mills} using the continuum limit of the molecular field theory.
Applying the procedure of Ref. \cite{Mills} to the case of the hcp(0001) surface, we obtain the
result that, for
\begin{equation}
1.5 - 2 \frac{J_b}{J_s} \geq 0 \; \mbox{or} \; \frac{J^s}{J^b} \geq 4/3
\label{2}
\end{equation}
there is in addition to the bulk Curie temperature $T^b_c$ a surface
Curie temperature $T^s_c$ which is connected to the bulk $T_c$ as:
\begin{equation}
T^s_c = [ 1 + (1.5 - 2 \frac{J_b}{J_s})^2 ] \; T^b_c \; ; \; T^b_c = \frac{12 J_b S (S+1)}{3 k_B}
\label{3}
\end{equation}
The ratio $\Delta E_{(\downarrow \uparrow - \uparrow \uparrow)}$/$\Delta E(AFM-FM)$ (cf., Table \ref{tab2})
is then used to determine $J^s/J^b$ in Eq.(2) \cite{isotropy}. 
In the case of an ideal surface the condition Eq.(2)
is not satisfied and there is no additional $T_c^s$. However, when the surface relaxation is taken into account,
the condition  Eq.(2) is satisfied and Eq.(3) yields $T_c^s \; = \; 1.33 \; T_c^b$ in very good
quantitative agreement with the recent experimental data \cite{Tober}
($T_c^s \; \approx \; 1.29 \; T_c^b$).

To summarize, we have presented the results of
one of the first applications of the LDA+U total energy method to study the
magnetic and electronic properties of a correlated metal.
We have found that the use of LDA+U instead of LDA yields
FM alignment between surface and bulk magnetic moments, in agreement with experiment.
An interlayer surface-to-bulk effective
exchange coupling is calculated to be close to its
bulk value for an ideal surface, but is enhanced by 90 \% by surface relaxation.
This enhancement is sufficiently strong to produce an elevated 
Curie temperature at the surface, as observed experimentally.
These results also have important implications for other rare earth (and possible also
transition metal) surfaces and interfaces, and suggest that both 4f (and 3d) correlation
and atomic structure must be accurately accounted for in order to quantitatively
describe magnetism.

This research was supported by National Science Foundation Grant DMR-9802076 and
U. S. Dept. of Energy, under Contract No. DE-AC03-76F00098.

\begin{table}[h]
\caption
{Spin magnetic moments ($M_s$ in $\mu_B$) for a Gd film with  
surface layer coupled 
parallel to the FM bulk 
resulting  from scalar-relativistic LDA+U calculations 
with experimental lattice constants.
For ``muffin-tin" (MT), these values are integrals over a ``muffin-tin'' sphere of radius 3.2 a.u.}
\begin{tabular}{ccccccc}
$M_s$ & layer & s   & p     & d     & f     & total \\
MT    & C     &0.016& 0.079 & 0.46  & 6.97  & 7.536 \\
MT    & S-2   &0.019& 0.081 & 0.49  & 6.97  & 7.576 \\
MT    & S-1   &0.011& 0.093 & 0.50  & 6.97  & 7.588 \\
MT    & S     &0.041& 0.074 & 0.67  & 6.975 & 7.773 \\
\hline
\multicolumn{2}{l}{Interstitial:} &2.076 &\multicolumn{2}{l}{Vacuum:}& 0.088 \\
\end{tabular}
\label{tab1}
\end{table}

\begin{table}[h]
\caption
{Total energy difference
between two magnetic configurations with the 
surface layer magnetically coupled
antiparallel ($\downarrow \uparrow$) and parallel ($\uparrow \uparrow$), 
$\Delta E_{(\downarrow \uparrow - \uparrow \uparrow)}$ (meV/atom), for
a Gd surface with (a) an ideal bulk atomic structure
(b) a relaxed structure from experiment and (c) an ``average'' (between bulk and relaxed surface)
values of surface-to-subsurface-to-bulk distances
and its ratio 
to the total energy difference between AFM and FM bulk ($\Delta E(AFM-FM)$ = 63 meV/atom):
$\Delta E_{(\downarrow \uparrow - \uparrow \uparrow)}$/$\Delta E(AFM-FM)$
} 
\begin{tabular}{cccc}
  & $\Delta E_{(\downarrow \uparrow - \uparrow \uparrow)}$&\multicolumn{2}{c}{
$\Delta E_{(\downarrow \uparrow - \uparrow \uparrow)}$/$\Delta E(AFM-FM)$} \\
a &  72   & & 1.14     \\
b & 135   & & 2.14    \\
c & 136   & & 2.16    \\
\end{tabular}
\label{tab2}
\end{table}
\end{multicols}

\newpage

\begin{multicols}{2}
\begin{center}
\begin{figure}
\label{ldad}
\psfig{file=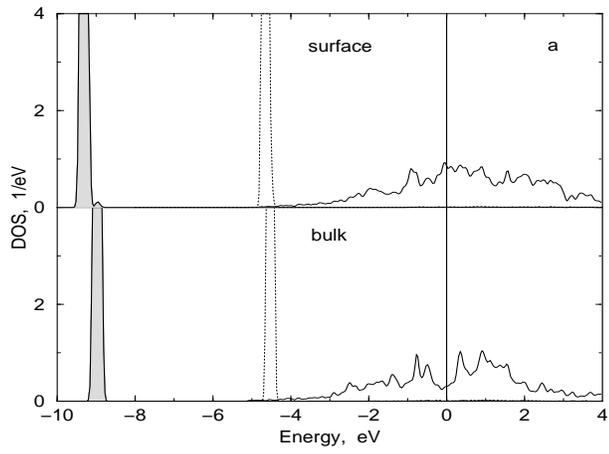,width=8cm,height=6cm}
\psfig{file=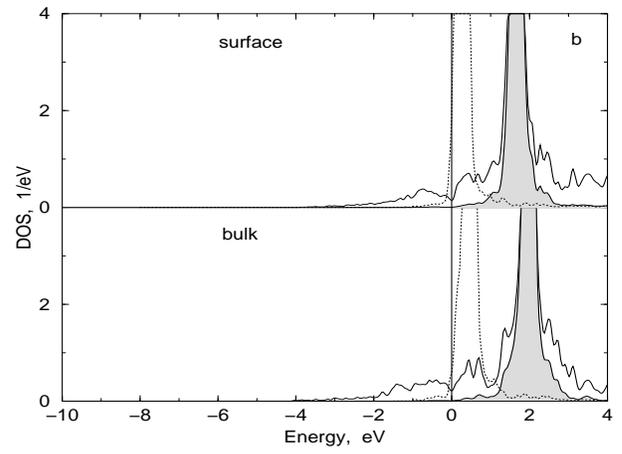,width=8cm,height=6cm}
\caption{DOS for Gd-film : LDA+U spin-up (a); spin-down (b), 4f-states (filled);
LDA 4f-states (dotted)}
\end{figure}
\end{center}
\end{multicols}

\begin{references}
\bibitem{symmetry}
The symmetry was chosen to be consistent with the
proper magnetic symmetry in the presence of spin-orbit coupling.
It allows us to do relativistic self-consistent
calculations starting from self-consistent scalar-relativistic
effective potential and exchange-correlation field.
\end{references}
\end{document}